\documentclass[aps,prb,amsmath,amssymb,floatfix,reprint,superscriptaddress]{revtex4-2}

\usepackage{graphicx} 
\usepackage{multirow}
\usepackage{amsfonts}
\usepackage{dsfont}
\usepackage{braket}
\usepackage{xcolor}
\usepackage{ulem}
\usepackage{comment}
\usepackage{natbib}
\usepackage{hyperref}
\hypersetup{colorlinks=true, linkcolor=blue, filecolor=blue,   citecolor=blue,   urlcolor=blue     }

\begin{document}

\preprint{APS/123-QED}

\title{Frustration effects on the magnetization plateau physics in a trimerized quantum spin-1/2 chain}

\author{L. M. Ramos}
\email{lucas.morais@ufms.br}
\affiliation{Instituto de Física, Universidade Federal de Mato Grosso do Sul, 79070-900, Campo Grande, MS, Brazil}

\author{M. Schmidt}
\email{mateus.schmidt@ufsm.br}
\affiliation{Departamento de Física, Universidade Federal de Santa Maria, 97105-900, Santa Maria, RS, Brazil}

\author{F. M. Zimmer}
\email{fabio.zimmer@ufms.br}
\affiliation{Instituto de Física, Universidade Federal de Mato Grosso do Sul, 79070-900, Campo Grande, MS, Brazil}

\begin{abstract}
We investigate frustration-induced instabilities in a trimerized quantum spin chain motivated by recent experimental findings for the compound Na$_2$Cu$_3$Ge$_4$O$_{12}$. 
Employing a cluster mean-field approach combined with Lanczos exact diagonalization, we analyze the ground-state and quantum-information properties of a Heisenberg model with competing interactions in a magnetic field.  
In the weakly frustrated regime, the system exhibits a robust $1/3$ magnetization plateau associated with a collective ferrimagnetic-like trimer state. 
Increasing the next-nearest-neighbor intratrimer coupling drives a pronounced reorganization of spin correlations, leading to a crossover toward a doublon-like correlation regime and providing a static ground-state picture consistent with the composite excitations observed dynamically in trimerized chains.
The resulting low-energy behavior can be interpreted in terms of weakly interacting emergent spins, offering a microscopic explanation for the extended stability of the magnetization plateau. 
Furthermore, through finite-size scaling analyses of the energy gap, von Neumann entanglement entropy, and fidelity susceptibility, we characterize the zero-field criticality of the model. 
Ultimately, our results suggest that frustration gives rise to qualitatively distinct quantum states and provide a microscopic framework for understanding the emergence of fractionalized excitations in trimerized quantum spin systems.

\end{abstract}

\maketitle

\section{INTRODUCTION}
One-dimensional (1D) and quasi-1D quantum systems constitute paradigmatic platforms for exploring fundamental aspects of many-body physics, as their reduced dimensionality dramatically enhances the role of quantum fluctuations \cite{giamarchi2003quantum,mikeska2004one}. 
Since the seminal works on exactly solvable models, such as the Ising, Heisenberg, and XY chains \cite{jordan1928paulische,bethe1931theorie,lieb1961two}, a wide variety of quantum phases and critical phenomena have been identified.
The presence of frustration, typically arising from competing exchange interactions, further enriches this landscape.
For instance, frustration famously suppresses conventional magnetic ordering, paving the way for exotic quantum states with no classical counterparts, including quantum spin liquids and topologically ordered phases \cite{anderson1987resonating,balents2010spin,savary2017quantum}.

Within this context, trimerized spin chains have attracted considerable attention due to their rich phase diagrams and distinct quantum critical behavior \cite{hida1994magnetic,okamoto1999magnetization,GuBO2006,Verkholyak2021}. 
Characterized by a periodic arrangement of coupled trimers, these structures provide an ideal testbed for investigating the subtle interplay between intra- and intertrimer correlations. 
A prominent experimental realization of such physics is found in the compound Na$_2$Cu$_3$Ge$_4$O$_{12}$, which features a distorted diamond-chain structure composed of Cu$^{2+}$ trimers \cite{yasui2014magnetic,chikara2023role,Han2024}. 
This material offers a unique window into the combined effects of frustration and strong quantum fluctuations. 
Recent experiments have revealed nontrivial fractionalized excitations, such as doublons and quartons, in its high-energy spectrum \cite{bera2022emergent,cheng2024quantum,li2025resonant,prabhakar2025fractionalized}. 
Furthermore, under an applied magnetic field, this compound exhibits a robust $1/3$ magnetization plateau, which is topologically protected by a finite thermodynamic energy gap \cite{kumar2025theoretical,sen2026trimerizedspin12chainemergent}.

Theoretical descriptions of this system typically rely on a spin-$1/2$ Heisenberg model featuring three distinct antiferromagnetic exchange couplings: the intratrimer interactions $J_1$ and $J_3$ and the intertrimer interaction $J_2$ \cite{cheng2024quantum, patnaik2025fermionic}. 
Microscopically, $J_{1}$ originates from a superexchange interaction mediated by oxygen ions, whereas $J_{2}$ arises from a super-superexchange path involving O–Ge–O bonds. 
In addition, $J_{3}$ represents a next-nearest-neighbor super-superexchange interaction connecting the two edge spins of a given trimer. 
Consequently, the competition between $J_{1}$ and $J_{3}$ acts as the primary source of frustration. 
In the weakly frustrated regime, the ground state is predominantly characterized by a highly entangled singlet configuration \cite{yasui2014magnetic}. 

Although the weakly frustrated regime of Na$_2$Cu$_3$Ge$_4$O$_{12}$ is relatively well understood \cite{bera2022emergent,cheng2024quantum}, the magnetic phase diagram in the strongly frustrated limit remains far less explored. In particular, the microscopic reorganization of ground-state correlations that may act as a static precursor to fractionalized composite excitations remains an open and actively debated theoretical problem \cite{kumar2025theoretical,patnaik2025fermionic}.
To address this challenge, we investigate the isotropic spin-$1/2$ Heisenberg model on a $J_1$-$J_2$-$J_3$ trimerized chain using a cluster mean-field (CMF) approach \cite{oguchi1955theory,yamamoto2009} combined with the Lanczos exact-diagonalization method. 
This methodology enables a rigorous treatment of strong local quantum correlations within clusters, while incorporating intercluster interactions self-consistently at the mean-field level. 
The CMF method has proven highly effective in investigating phase transitions across a diverse array of magnetic systems. 
Notable applications include the classical Ising model on square \cite{jin2013phase,godoy2020ising}, triangular \cite{malakar2020phases}, and kagome \cite{schmidt2017spin} lattices, as well as the Ising model in the presence of quantum fluctuations \cite{kellermann2019quantum,zimmer2016quantum} and the Heisenberg model with competing interactions \cite{WIESER2021168414,RAMOS2026173687}. 
Furthermore, the CMF approach has been successfully employed to explore frustrated magnets under applied magnetic fields, yielding theoretical predictions that qualitatively reproduce experimental observations for complex compounds such as $\mathrm{CuInVO}_5$ \cite{Singhania}, $\mathrm{TmMgGaO}_4$ \cite{PhysRevResearch.2.043013}, Ba$_3$CoSb$_2$O$_9$ \cite{PhysRevLett.114.027201},
and ($o$-MePy-V)PF$_6$ \cite{ramos2025interplay}.

By systematically varying the frustration ratio $J_3/J_1$, we map out the effects of the external magnetic field and frustration on the ground state. 
Furthermore, to accurately identify quantum phase boundaries and characterize the nature of the underlying critical points, we employ quantum information measures, namely the fidelity susceptibility \cite{zanardi2006,GU2008,GU2010} and the von Neumann entanglement entropy \cite{Pasquale2004}. 
Together with finite-size scaling analyses, these tools provide valuable insights into the frustration-driven criticality of the model.

The remainder of this paper is organized as follows. In Sec.~\ref{sec:model}, we introduce the theoretical model and detail the cluster mean-field framework combined with exact diagonalization. Section~\ref{sec:results} is devoted to our numerical results, where we characterize the ground-state from the discussion of the influence of strong frustration on the phase diagram and quantum information measures. 
Finally, Sec.~\ref{sec:summary} summarizes our main findings and contains our conclusions.

\section{MODEL AND METHOD}\label{sec:model}
\begin{figure}[t]
    \centering
    \includegraphics[width=1.0\linewidth]{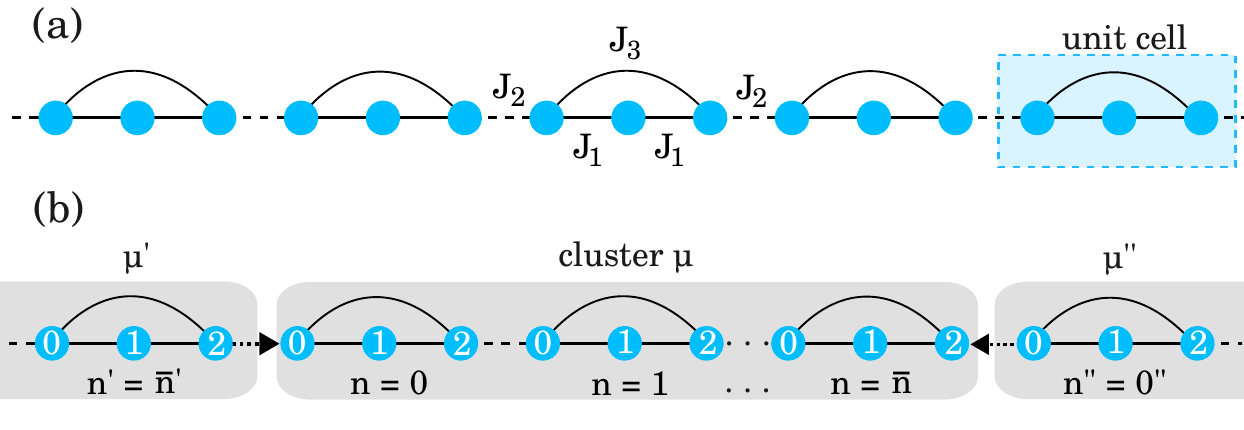}
    \caption{(a) Schematic representation of the trimer quantum spin chain with nearest-neighbor exchange interactions $J_1$, $J_2$, and a next-nearest-neighbor exchange interaction $J_3$. The shaded blue region denotes the unit cell of the model. (b) Schematic representation of the CMF approach applied to a trimerized quantum spin chain, in which the infinite chain is partitioned into identical clusters (shaded gray region) of $L_c$ sites. The $\bar{n}$ represents the last trimer of the central cluster that is equal to $\bar{n} = (L_c-3)/3$.}
    \label{fig:f1}
\end{figure}

We consider the isotropic spin-$1/2$ Heisenberg model on a trimerized chain, schematically depicted in Fig.~\ref{fig:f1}(a). 
The system is governed by the Hamiltonian
\begin{equation}\begin{split}\label{eq:hamiltonian}
H = \sum_{n=0}^{L/3-1} [ &J_1 (\mathbf{S}_{n,0}\cdot\mathbf{S}_{n,1}+\mathbf{S}_{n,1}\cdot\mathbf{S}_{n,2}) + J_2(\mathbf{S}_{n,2}\cdot\mathbf{S}_{n+1,0})\\
   &+J_3(\mathbf{S}_{n,0}\cdot\mathbf{S}_{n,2})] - h^{z}\sum_{n,\alpha}S^{z}_{n,\alpha},
\end{split}\end{equation}
where $\mathbf{S}_{n,\alpha}$ denotes the spin-$1/2$ operator at unit cell $n$ and sublattice index $\alpha \in \{0,1,2\}$, for a chain of $L$ spins.
The parameters $J_{1}$ and $J_{3}$ represent the AF nearest-neighbor and next-nearest-neighbor intratrimer exchange interactions, respectively, while $J_{2}$ denotes the AF intertrimer coupling. 
The external longitudinal magnetic field is given by $h^{z}$. 
Frustration arises primarily from the competition between the AF interactions $J_{1}$ and $J_{3}$ within each trimer unit. 

To investigate the ground-state properties in the thermodynamic limit ($L\rightarrow\infty$), we employ the CMF theory. 
In this approach, the infinite chain is partitioned into identical clusters of $L_c$ sites. 
The intracluster interactions are treated exactly via diagonalization, whereas the intercluster interactions are decoupled at the mean-field level.
Specifically, the interaction between a spin $\mathbf{S}_i$ at the cluster boundary and a spin $\mathbf{S}_j$ in the neighboring cluster is approximated as
\begin{equation}\label{eq:mean}
\mathbf{S}_i \cdot \mathbf{S}_j \approx \mathbf{S}_i \cdot \langle \mathbf{S}_j \rangle + \langle \mathbf{S}_i \rangle \cdot \mathbf{S}_j - \langle \mathbf{S}_i \rangle \cdot \langle \mathbf{S}_j \rangle,
\end{equation}
where $\langle \mathbf{S} \rangle$ represents the ground-state expectation value of the local magnetization. 
This decoupling yields the Hamiltonian 
\begin{equation}\label{eq:cmf}
H= \sum_{\mu}\left[H_{\text{exact}}^{\mu}+H_{\text{MF}}\right],  
\end{equation}
where $H_{\text{exact}}^{\mu}$ contains the exact interactions of the set of trimers within each cluster $\mu$ and $H_{\text{MF}}$ represents the boundary term consisting of effective mean fields acting on the edge spins of the cluster (see Fig.~\ref{fig:f1}(b)), expressed as

\begin{equation}\begin{split}\label{eq:boundary}
H_{\text{MF}} =  J_{2}[\mathbf{S}_{0,0} \cdot \langle \mathbf{S}_{\bar{n}',2} \rangle 
+ \mathbf{S}_{\bar{n},2}  \cdot \langle \mathbf{S}_{0'',0}\rangle\\
-1/2 (\langle \mathbf{S}_{\bar{n}',2} \rangle \cdot \langle \mathbf{S}_{0,0} \rangle+\langle \mathbf{S}_{\bar{n},2} \rangle \cdot \langle \mathbf{S}_{0'',0} \rangle)].
\end{split}\end{equation}
where $\bar{n}$ represents the last trimer of the cluster, while the prime notation specifies trimers belonging to neighboring clusters. The factor 1/2 was inserted to avoid double-counting in the mean fields.

We adopt translational invariance on the mean-field parameters, ensuring that the effective fields acting on the cluster edges correspond exactly to the local magnetizations of the equivalent sites within the cluster, i.e.,

\begin{equation}
\langle \mathbf{S}_{j} \rangle_{\mu} \equiv \langle \mathbf{S}_{j}\rangle_{\mu'(\mu'')}.
\end{equation}
Based on this assumption, the effective single-cluster Hamiltonian becomes
\begin{equation}\begin{split}
    H_{\text{eff}}=H_{\text{intra}} &+ J_{2}[\mathbf{S}_{\bar{n},2} \cdot \langle \mathbf{S}_{0,0}\rangle +\langle \mathbf{S}_{\bar{n},2}\rangle\cdot\mathbf{S}_{0,0}\\&-\langle \mathbf{S}_{\bar{n},2}\rangle\cdot\langle\mathbf{S}_{0,0}\rangle],
\end{split}\end{equation}
where $H_{\text{intra}}$ is given by Eq. \ref{eq:hamiltonian} with $L=L_c$ without boundary conditions. 
The $H_{\text{eff}}$ is solved self-consistently, in which the ground-state wave function $|\psi_{0}\rangle$ is obtained using the Lanczos algorithm \cite{lanczos1950iteration}. 
To preserve numerical stability and eliminate ghost states, which are particularly problematic in the strong-frustration regime, we also implement a re-orthogonalization scheme \cite{parlett1998symmetric}. 
The local magnetizations are then updated according to
\begin{equation}\label{eq:local_mag}
\mathbf{m}_i = \langle \psi_{0}| \mathbf{S}_{i} |\psi_{0}\rangle,
\end{equation}
until the convergence criterion is satisfied. 

Once self-consistency is achieved, the quantum phases are characterized through several observables. 
The local magnetization profile provides evidence of magnetic ordering and plateau formation in the total magnetization per spin
\begin{equation}
    M=\sqrt{\sum (\langle S_{n,i}^x \rangle^2+\langle S_{n,i}^y \rangle^2+\langle S_{n,i}^z \rangle^2)}/L_c,
\end{equation}
while the two-point spin-spin correlation function,
\begin{equation}
C_{ij}(r) = \langle \mathbf{S}_{n,i} \cdot \mathbf{S}_{n+r,j} \rangle,
\end{equation}
provides insights into the microscopic structure of the ground state and the spatial organization of magnetic correlations. 
The indices $i,j=0,1,2$ identify the sites inside the trimer $n$, and the distance between two trimers $r$ (e.g., $r=0$ or 1) denotes the intratrimer (intertrimer) correlations.

Ultimately, the CMF approach adopted here establishes an ideal framework for capturing the interplay between strong local quantum correlations and the collective effects induced by frustration. 
By treating intracluster interactions exactly while incorporating intercluster couplings at the mean-field level, this method allows for a consistent and highly accurate description of microscopic observables.

To further probe underlying quantum criticality, we also perform exact diagonalization calculations for finite clusters, without the mean-field decoupling, to carry out finite-size scaling analyses. 
In this case, we analyze the fidelity susceptibility $\chi_F$, defined as \cite{GU2008,GU2010}
\begin{equation}\chi_F(\lambda) = \lim_{\delta\lambda\rightarrow0} \frac{-2\ln F(\lambda,\delta\lambda)}{(\delta\lambda)^{2}} \approx \frac{2\left[1 - F(\lambda, \lambda + \delta\lambda)\right]}{(\delta\lambda)^2},
\end{equation}
where $F(\lambda, \lambda + \delta\lambda) = |\langle \psi_0(\lambda) | \psi_0(\lambda + \delta\lambda) \rangle|$ is the fidelity between ground states separated by a small parameter shift $\delta\lambda$. 
In our study, the driving parameter $\lambda$ corresponds to either the frustration ratio $J_3/J_1$ or the applied magnetic field $h^{z}/J_{1}$, with the shift magnitude set to $\delta\lambda=10^{-4}$. 

Complementarily, we compute the von Neumann entanglement entropy,
\begin{equation}S_{vN} = -\text{Tr}(\rho_A\ln\rho_A),
\end{equation}
where $\rho_A = \text{Tr}_{\text{B}}|\psi_0\rangle\langle\psi_0|$ is the reduced density matrix of a subsystem $A$ of length $L_A$, obtained by tracing out the environmental degrees of freedom $B$. For a critical gapless system described by a conformal field theory (CFT), the entropy scales with the subsystem size as \cite{Pasquale2004}
\begin{equation}\label{eq:vn_scaling}S_{vN}(L_c,L_{A}) = \frac{c}{3\eta}\ln\left[ \frac{\eta L_c}{\pi}\sin \left(\frac{\pi L_{A}}{L_c}\right) \right]+a_{\eta},
\end{equation}
where $c$ is the central charge, $\eta=1$ ($\eta=2$) for periodic (open) boundary conditions, and $a_\eta$ is a non-universal constant. 
The entanglement entropy provides a powerful, unbiased metric for identifying the universality class of one-dimensional quantum systems. 
This approach is particularly advantageous because it relies exclusively on ground-state properties, thereby circumventing explicit knowledge of the full excitation spectrum or non-universal parameters such as the sound velocity \cite{calabrese2009}. 

In the following section, we present a detailed analysis of the magnetic response, correlation patterns, and quantum-information measures characterizing the ground-state phase diagram. 

\section{RESULTS AND DISCUSSION}\label{sec:results}

The set of self-consistent equations, defined by the cluster-edge local magnetizations (Eq. (\ref{eq:local_mag})) and the effective Hamiltonian (Eq. (\ref{eq:cmf})), is solved iteratively for clusters of up to $L_{c}=24$ sites.
The effective cluster Hamiltonian is diagonalized using the Lanczos algorithm. 
In our numerical calculations, we set the energy scale by fixing $ J_1$ to unity.
We treat the intratrimer ratio $J_3/J_1$ as a control parameter for the degree of frustration, and we adjust the intertrimer coupling to $J_2/J_1=0.18$, which is consistent with density matrix renormalization group (DMRG) estimates and experimental measurements for the compound Na$_2$Cu$_3$Ge$_4$O$_{12}$ \cite{bera2022emergent}.
This allows us to map the magnetic phase diagram and explore the regimes of weak ($J_3/J_1$ away from 1) and strong ($J_3/J_1 \rightarrow 1$) frustration.
The quantum phases are characterized by analyzing the ground-state energy, energy gap, local magnetization profiles, and spin-spin correlation functions. 
Furthermore, to identify quantum phase transitions with high precision, we compute quantum information measures, specifically the fidelity susceptibility and the von Neumann entanglement entropy. 
We begin by presenting the effect of the magnetic field on the phase diagram, followed by an analysis of the quantum criticality of the system at zero magnetic field.

\subsection{The magnetic ground-state}
\begin{figure}[!t]
    \centering
    \includegraphics[width=0.95\linewidth]{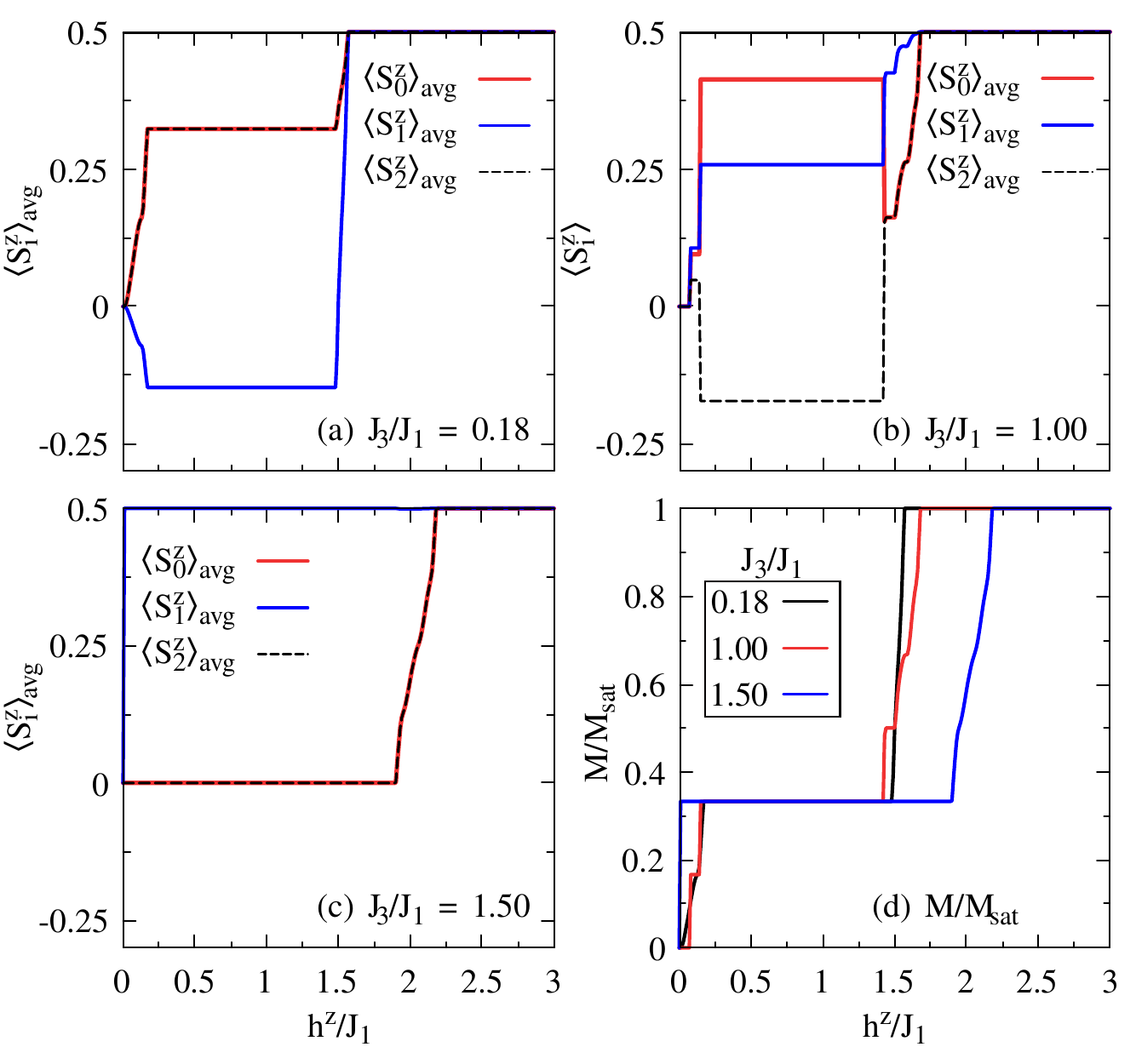}
    \caption{ Magnetic ground state of the trimerized quantum spin chain obtained from the 12-site CMF method. 
    (a) Longitudinal components of the trimer for $J_3/J_1=0.18$, (b) Longitudinal components of the trimer for $J_3/J_1=1.00$, and (c) Longitudinal components of the trimer for $J_3/J_1=1.50$. (d) Total magnetization normalized by the saturation for different $J_3/J_1$. To minimize the boundary effects of the mean fields, we perform a spatial averaging of the local magnetizations over the equivalent trimers within the 12-site CMF approach.}
    \label{fig:mags}
\end{figure}

An interesting fact about the compound Na$_2$Cu$_3$Ge$_4$O$_{12}$ is that it represents a well-characterized experimental realization of a quantum one-dimensional system and is described by a trimerized spin-chain model with three distinct antiferromagnetic exchange interactions, $J_1$, $J_2$, and $J_3$ \cite{yasui2014magnetic}.
Motivated by the experimental findings of this compound, we investigate the site-resolved local magnetizations, $\langle S_i^z \rangle$, and the total magnetization $M/M_{\textrm{sat}}$ as a function of the external field $h^z$ by varying frustration ratios $J_3/J_1$ for $J_2/J_1 = 0.18$ (see Fig. \ref{fig:mags}).

To minimize the boundary effects of the mean fields, we perform a spatial averaging of the local magnetizations over the equivalent trimers within the 12-site CMF approach.
In the weak-frustration regime ($J_3/J_1 = 0.18$), the system undergoes a transition from a nonmagnetic ground state ($\langle S_i^z \rangle_{\textrm{avg}}=0$) to a polarized state ($\langle S_i^z \rangle_{\textrm{avg}}=1/2$), passing through an intermediate phase characterized by plateaus in the local magnetization (see Fig. \ref{fig:mags}(a)).
Notably, within the field range $0.15\lesssim h^z/J_1 \lesssim 1.50$, the local moments display a clear ferrimagnetic-like structure, where the central spin (see the blue curve) tends to align antiparallel to the external field, while the edge spins (red and black curves) tend to align parallel.
This local magnetic behavior results in the emergence of a large $1/3$ magnetization plateau state, as shown in Fig. \ref{fig:mags}(d) (see the black curve).
Importantly, the present cluster mean-field scheme provides critical fields delimiting the plateau phase and the saturation in agreement with previous DMRG results for Na$_2$Cu$_3$Ge$_4$O$_{12}$ \cite{kumar2025theoretical}. 

The previously discussed behavior of the local magnetizations persists to strengths of $J_3$ slightly below $J_1$. 
However, as the next-nearest-neighbor interaction increases to $J_3/J_1 = 1.0$ (Fig. \ref{fig:mags}(b)), the ferrimagnetic spin state changes, becoming characterized by a tendency of {\it up-up-down} spin-like structures within the trimers.
A further increase in $J_3/J_1$ drives the system into a significant reorganization of the ground state, as shown in Fig. \ref{fig:mags}(c) for $J_3/J_1 = 1.5$. 
The local moments of the edge spins in trimers vanish ($\langle S_{\textrm{edge}}^z \rangle \approx 0$) over a wide field range, while the central site contains a nearly free spin exhibiting a field-driven paramagnetic-like behavior. 
In this case, the 1/3 magnetization plateau state covers practically the entire field range before saturation (see the blue curve in Fig. \ref{fig:mags}(d)).
This behavior indicates a decoupling of the trimer degrees of freedom for $J_3>J_{1}$: the strong next-nearest-neighbor interactions effectively lock the edge spins into a nonmagnetic state, leaving the central spin to respond independently to the field. 
The local magnetization results indicate a crossover for $J_3/J_1\approx 1$, signaling the onset of a partially disordered phase driven by the competition between the intratrimer interactions $J_1$ and $J_3$.

\begin{figure}[!t]
    \centering
    \includegraphics[width=0.95\linewidth]{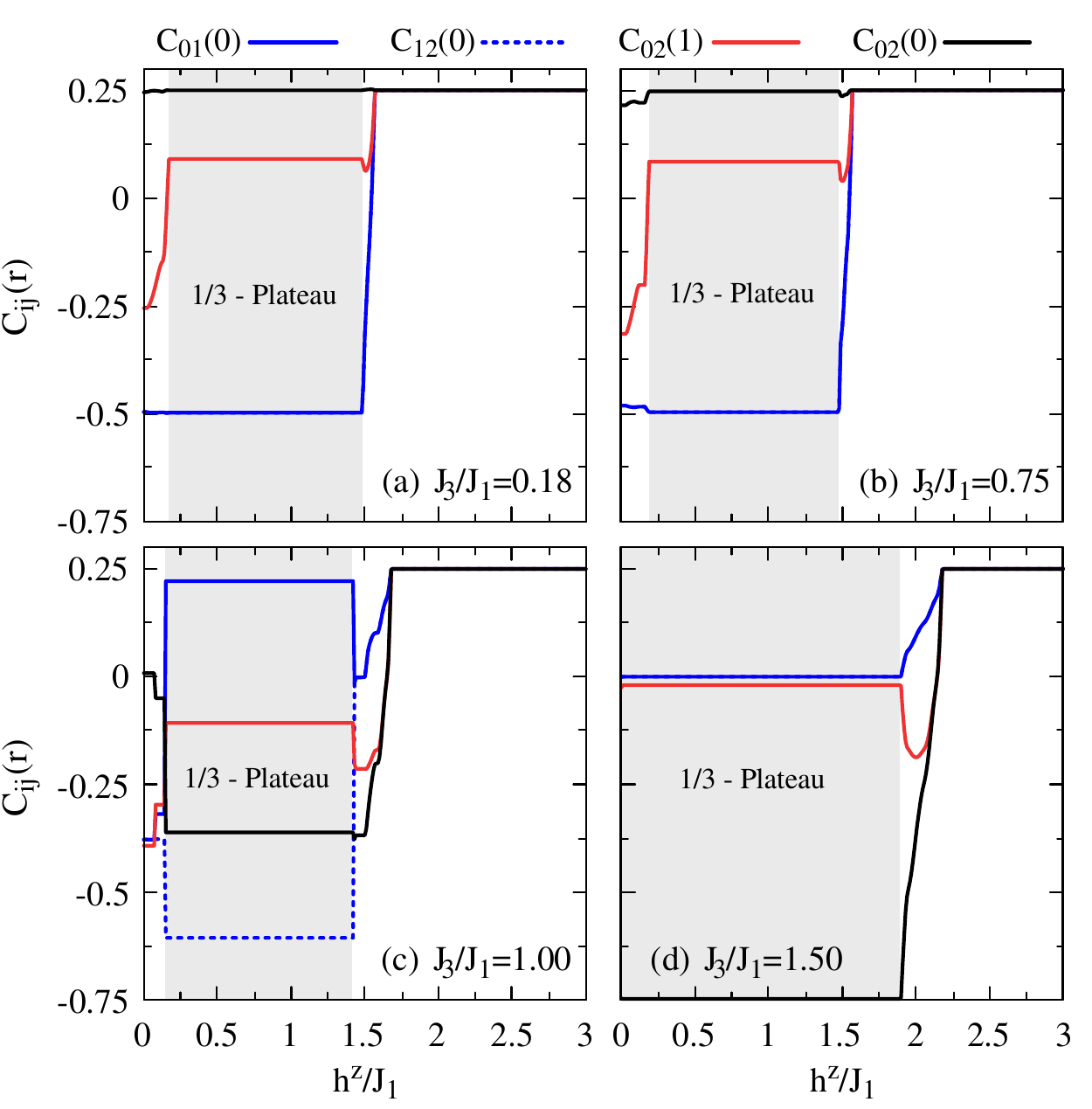}
    \caption{Field dependence of the spin–spin correlation functions computed within the 12-site CMF approach at zero temperature ($T/J_{1}=0$) for different values of the frustration parameter: (a) $J_{3}/J_{1}=0.18$, (b) $J_{3}/J_{1}=0.75$, (c) $J_{3}/J_{1}=1.00$, and (d) $J_{3}/J_{1}=1.50$.
    In panels (a), (b) and (d) the correlations $C_{01}(0)$ and $C_{12}(0)$ are equivalent. To minimize boundary effects arising from the mean-field treatment, all correlations were computed by averaging over all trimers of the cluster.}
    \label{fig:correlations_vs_hz}
\end{figure}

To further extend the microscopic picture, we present the magnetic-field dependence of two-point spin–spin correlation functions in Fig. \ref{fig:correlations_vs_hz}. 
In the weakly frustrated case (Fig. \ref{fig:correlations_vs_hz}(a) for $J_3/J_1=0.18$), the correlations display distinct features that evolve differently as the magnetic field increases. 
The nearest-neighbor intratrimer correlations $C_{01}(0) \approx C_{12}(0)$ exhibit a strong AF character, converging to approximately $-0.50$.
This value corresponds to the theoretical expectation for the internal bonds of an isolated spin-$1/2$ trimer in its doublet ground state \cite{patnaik2025fermionic}.
Meanwhile, the next-nearest-neighbor intratrimer correlation $C_{02}(0)$ remains finite and FM.
This behavior is consistent with the ferrimagnetic structure identified in the local magnetization profiles and reflects a collective trimer ground state rather than the formation of isolated singlet pairs.
Furthermore, the intertrimer correlations $C_{02}(1)$ remain small within the plateau region, indicating a weak coupling between trimers.
These spin–spin correlations follow the same trend as $J_3/J_1$ increases up to 0.75 (see Fig. \ref{fig:correlations_vs_hz}(b)). 
Particularly, these are well-established results from recent works \cite{bera2022emergent,cheng2024quantum,Han2024}, in which the zero-field ground-state wave function is a linear combination of trimers, each dominated by a singlet-like configuration on the bond $J_{1}$. 

However, at  $J_3/J_1=1.00$, an interesting reorganization of the correlation pattern is observed (see Fig. \ref{fig:correlations_vs_hz}(c)). 
In low magnetic fields, $C_{02}(0)$ approaches zero, while the intertrimer correlation $C_{02}(1)$ becomes the dominant AF one, signaling the relevance of intertrimer coupling in this frustrated regime. 
When the field leads to the magnetization plateau region,  $C_{02}(0)$ correlation becomes AF with the nearest-neighbor correlations $C_{01}(0)$ and $C_{12}(0)$ being consistent with the {\it up-up-down} structure previously discussed.
It is worth noting that $C_{02}(0)$ changes from FM to AF with increasing frustration (as $J_3/J_1$ increases from 0.18 to 1.00).
This reconfiguration of correlations provides microscopic evidence of the relevance of the frustration in this system.

Another scenario arises as $J_3/J_1$ is further increased (see Fig. \ref{fig:correlations_vs_hz}(d) for $J_3/J_1=1.50$). 
In this case, the correlation $C_{02}(0)$ remains pinned at $-0.75$ over a broad field range ($h^z/J_1 \lesssim 2.0$), corresponding to the expectation value of a perfect singlet formed by two spin-$1/2$ moments located at the trimer edges.
This indicates that the edge spins of the trimer form a nonmagnetic singlet that is largely insensitive to the applied magnetic field.
Concomitantly, the correlations with the central spin, such as $C_{01}(0)$ and $C_{12}(0)$, remain strongly suppressed, signaling an effective decoupling of the central site from the singlet pair. 
This decoupling explains the vanishing edge local magnetizations observed in Fig. \ref{fig:mags}(c). In addition, it establishes a microscopic picture in which the dominance of $J_3$ interaction drives the system into a dimerized quantum state of the trimer-edge spins. 
In contrast, the central spin behaves as a field-polarized degree of freedom.
These findings suggest that frustration induces a fundamental reorganization of intratrimer correlations, leading to a quantum phase transition from a collective trimer state to a phase characterized by static singlet-pair formation. 

The above interpretation provides a complementary perspective to the emergent composite-excitation picture established experimentally and theoretically for Na$_2$Cu$_3$Ge$4$O${12}$ in Refs.~\cite{bera2022emergent,cheng2024quantum}. 
In these works, doublons and quartons were identified as high-energy composite quasiparticles associated with internal trimer degrees of freedom, with their dispersions and spectral weights resolved through inelastic neutron scattering and DMRG calculations. 
Our results provide a direct microscopic connection between the doublon excitation and the underlying ground-state correlations in different frustrated regimes. 
In particular, the pinning of the intratrimer correlation $C_{02}(0)$ to the singlet value $-0.75$ and the suppression of correlations involving the central spin suggest that the doublon identified dynamically in Refs.~\cite{bera2022emergent,cheng2024quantum} has a static precursor in the ground-state correlation pattern. 
Within this regime, the low-energy behavior can be interpreted as an inert doublon weakly coupled to an emergent spin-1/2 degree of freedom, providing a natural effective description of the system. 
On the other hand, quartons can be viewed as higher-order composite excitations associated with the internal excited states of the trimers, whose propagation and eventual fractionalization are beyond the static description provided by the present CMF approach.

\begin{figure}[!t]
    \centering
    \includegraphics[width=0.85\linewidth]{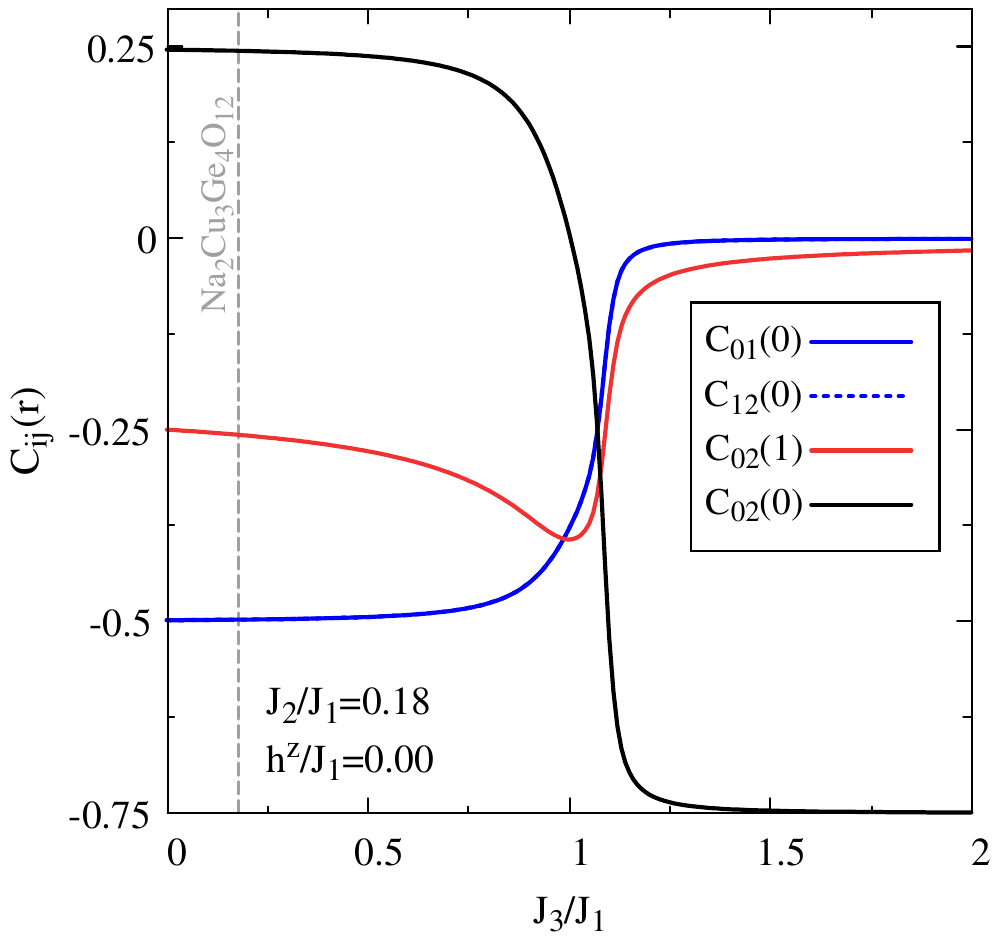}
    \caption{Evolution of the spin–spin correlation functions at zero magnetic-field ($h^{z}/J_{1}=0$) as a function of the frustration ratio $J_3/J_1$ for $J_2/J_1=0.18$ and periodic boundary conditions. The gray dashed line indicates the parameter range relevant to Na$_2$Cu$_3$Ge$_4$O${_{12}}$.}
    \label{fig:correlations_vs_J3}
\end{figure}

Figure \ref{fig:correlations_vs_J3} provides further microscopic insight into the ground-state reorganization, showing the dependence of the spin–spin correlation functions on the frustration ratio $J_3/J_1$ for $J_2/J_1=0.18$ and zero magnetic field.
In this region, an increase in $J_3/J_1$ drives the next-nearest-neighbor correlation $C_{02}(0)$ from an FM character to the exact singlet value of $-0.75$, whereas the nearest-neighbor correlations $C_{01}(0)$ and $C_{12}(0)$ change from an AF-type to an uncorrelated regime.
These behaviors reflect the tendency of the edge spins to bind into a perfect singlet, consistent with the doublon-like picture identified in the $J_3$ dominance regime.
This correlation-driven mechanism provides a unifying microscopic explanation for the magnetization patterns, plateau stability, and emergent composite excitations discussed throughout this work.
Furthermore, these results indicate the presence of a crossover region marking a transition from a collective trimerized state to a dimerized state at the static level.

\begin{figure}[!t]
    \centering
    \includegraphics[width=1.0\linewidth]{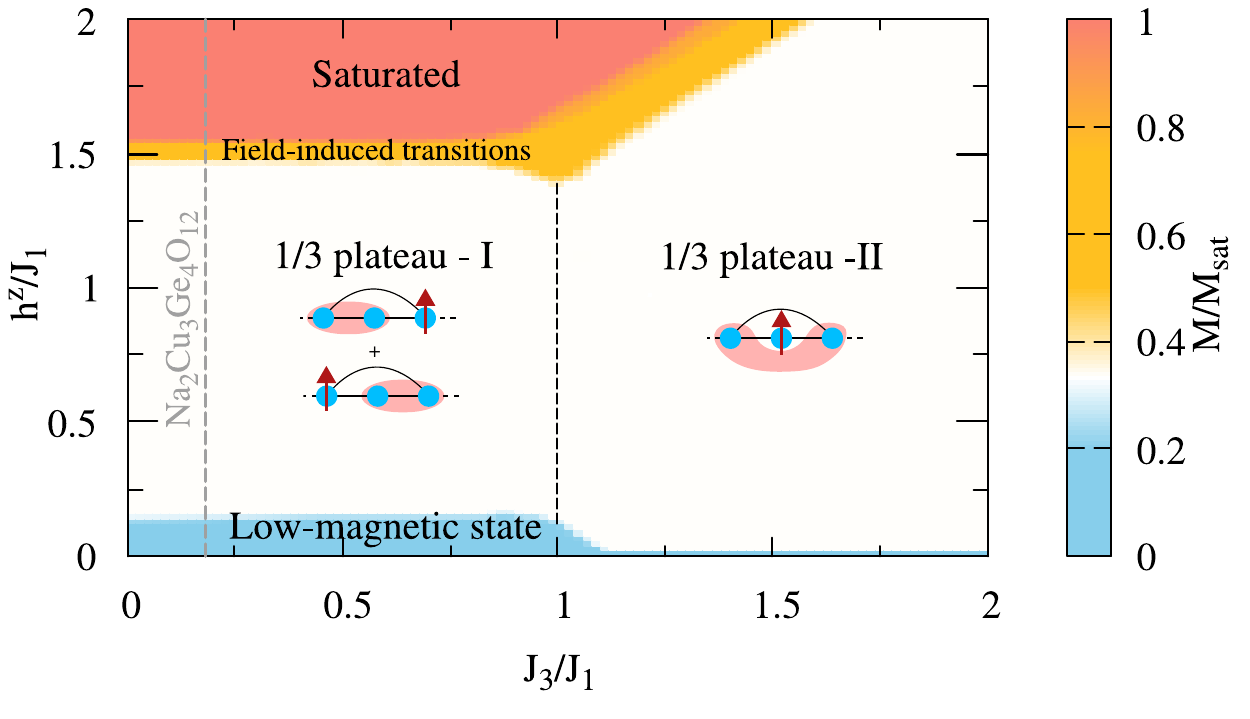}
    \caption{Magnetic phase diagram in the plane $h^{z}/J_{1}-J_3/J_1$ obtained within the CMF framework for $J_2/J_1=0.18$ and $L_{c}=12$ sites. Distinct magnetic regimes are identified, including the low-magnetization state, two distinct 1/3 magnetization plateau phases, a field-induced phase, and the fully polarized state. The black dashed lines denote the crossover region separating the trimerized ground state from the static doublon-like phase. The gray dashed line indicates the parameter range relevant to Na$_2$Cu$_3$Ge$_4$O${_{12}}$.}
    \label{fig:phase_diagram}
\end{figure}
The magnetic phase diagram in Fig. \ref{fig:phase_diagram} summarizes our findings of this section. 
Several distinct magnetic states can be clearly identified, reflecting the interplay between frustration and the external magnetic field. 
In the low-frustration region ($J_3/J_1\lesssim1$), the system exhibits a low-magnetization ground state at weak fields, followed by a robust 1/3 magnetization plateau (denoted by 1/3-Plateau I). 
This plateau is associated with the tendency toward the {\it up-down-up} ferrimagnetic-like state within each trimer, as previously inferred from the local magnetization profiles and correlation functions. 
The plateau width remains substantial over a broad range of fields, indicating the stability of this partially polarized state. 
As the frustration parameter approaches $J_3/J_1\approx1$, the phase boundaries exhibit pronounced variations. 
This regime coincides with the anomalies observed in the correlation functions and signals a crossover region that is enhanced by frustration. 
The strong sensitivity of the phase diagram near this region reflects the competition between intratrimer interactions, which can drive the system toward a quantum critical instability. 
In the regime $J_3/J_1>1$, the magnetic response undergoes a qualitative change. 
The plateau structure persists but evolves into a distinct state (denoted by 1/3 - Plateau II), consistent with the formation of an effective dimerized state and partially decoupled spins. 
This behavior aligns with the correlation analysis, which revealed the emergence of perfect singlets within the trimers. 
It means that the plateau under a strong $J_3/J_1$ regime presents a microscopic origin that differs from the weak-frustration regime. 
At sufficiently large magnetic fields, the system transitions to the fully polarized phase. 
The field-induced transition boundary to saturation (yellowish region) shows only a weak dependence on $J_3/J_1$, indicating that the high-field behavior is primarily governed by Zeeman energy. 

In summary, our results for the magnetic phase diagram corroborate and extend those reported by Bera {\it{et al.}} in Ref. \cite{bera2022emergent} and Cheng {\it{et al.}} in Ref. \cite{cheng2024quantum}.
In our case, we find that the ground state of the trimerized $J_1$-$J_2$-$J_3$ quantum spin chain develops static correlations consistent with a doublon-like state.
Furthermore, these findings highlight the rich interplay between frustration, quantum fluctuations, and field-induced effects, providing a unified framework for understanding the unconventional magnetic behavior observed in trimerized quantum spin systems.

In the following section, we investigate how the frustration parameter $J_3/J_1$ modifies the finite-size behavior in the sector of $M=0$ around the crossover region of the trimerized quantum spin chain. This analysis provides an ideal framework for exploring the critical features and quantum information properties of the frustrated model in the zero-field limit. In the absence of an external magnetic field, the self-consistent local magnetizations vanish due to $SU(2)$ symmetry, effectively reducing the cluster mean-field Hamiltonian to that of an isolated finite cluster. Consequently, periodic boundary conditions are employed to mitigate finite-size boundary effects and accurately capture the bulk properties.
Particularly, the periodic boundary condition was already adopted for obtaining the results discussed in Fig. (\ref{fig:correlations_vs_J3}).
\subsection{Zero-field quantum criticality and finite-size scaling}

\begin{figure}[!t]
    \centering
    \includegraphics[width=1.0\linewidth]{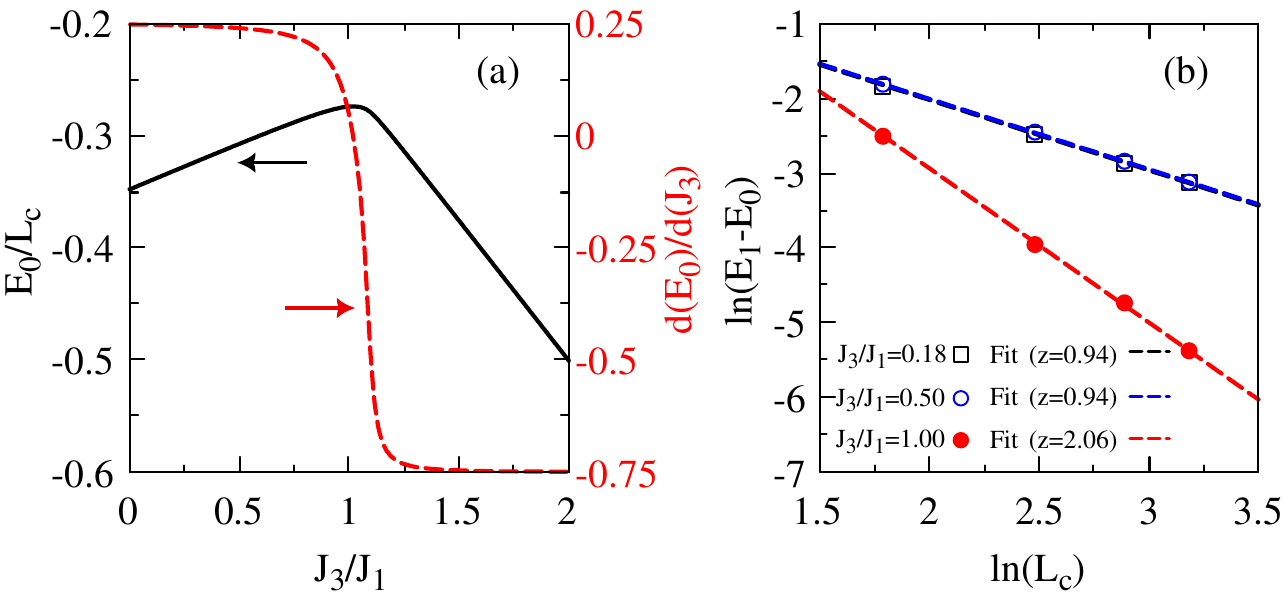}
    \caption{ (a) Ground-state energy per site $E_0/L_c$ (left axis), and its derivative (right axis), showing a smooth evolution across the parameter range for $J_2/J_1=0.18$, $h^z/J_1=0.0$, and $L_{c}=12$ sites. (b) Log-log behavior of energy gap $(E_1-E_0)$ as a function of $L_{c}$ for three different frustration ratios $J_3/J_1=0.18$, $J_3/J_1=0.50$ and $J_3/J_1=1.00$. The dashed lines represent the linear fit of the data.}
    \label{fig:gs_gap}
\end{figure}

Further insight into the frustration-driven instability is obtained from the behavior of the ground-state energy and the finite-size scaling of the energy gap, presented in Fig.~\ref{fig:gs_gap}. 
The ground-state energy per site and its derivative vary smoothly over the entire range of the frustration parameter $J_3/J_1$, as shown in Fig.~\ref{fig:gs_gap}(a). 
In particular, the ground state energy reaches a maximum value at $J_{3}/J_1\approx 1.02$, where its derivative vanishes. By virtue of the Hellmann-Feynman theorem \cite{hellman1937einfuhrung,Feynman1939},
\begin{equation}
    \frac{\partial E_{n}}{\partial\lambda}=\langle \psi_{n}|\frac{\partial H}{\partial\lambda}|\psi_{n}\rangle,
\end{equation}
setting $\lambda = J_3/J_1$ implies that this derivative of the ground-state energy corresponds exactly to the next-nearest-neighbor intratrimer correlation function $C_{02}(0)$ depicted in Fig.~\ref{fig:correlations_vs_J3}. 
Consequently, this energy extremum marks the point where $C_{02}(0)$ vanishes, changing its character from ferromagnetic to antiferromagnetic.
Furthermore, the absence of visible discontinuities in the first derivative supports the interpretation that the transition is not driven by a first-order level crossing \cite{CAROLLO20201}. 
Instead, this smooth behavior is characteristic of a crossover regime, where frustration progressively reshapes the dominant local quantum fluctuations and redistributes the internal correlations without inducing abrupt jumps in the physical quantities.
\begin{figure}[!t]
    \centering
    \includegraphics[width=0.9\linewidth]{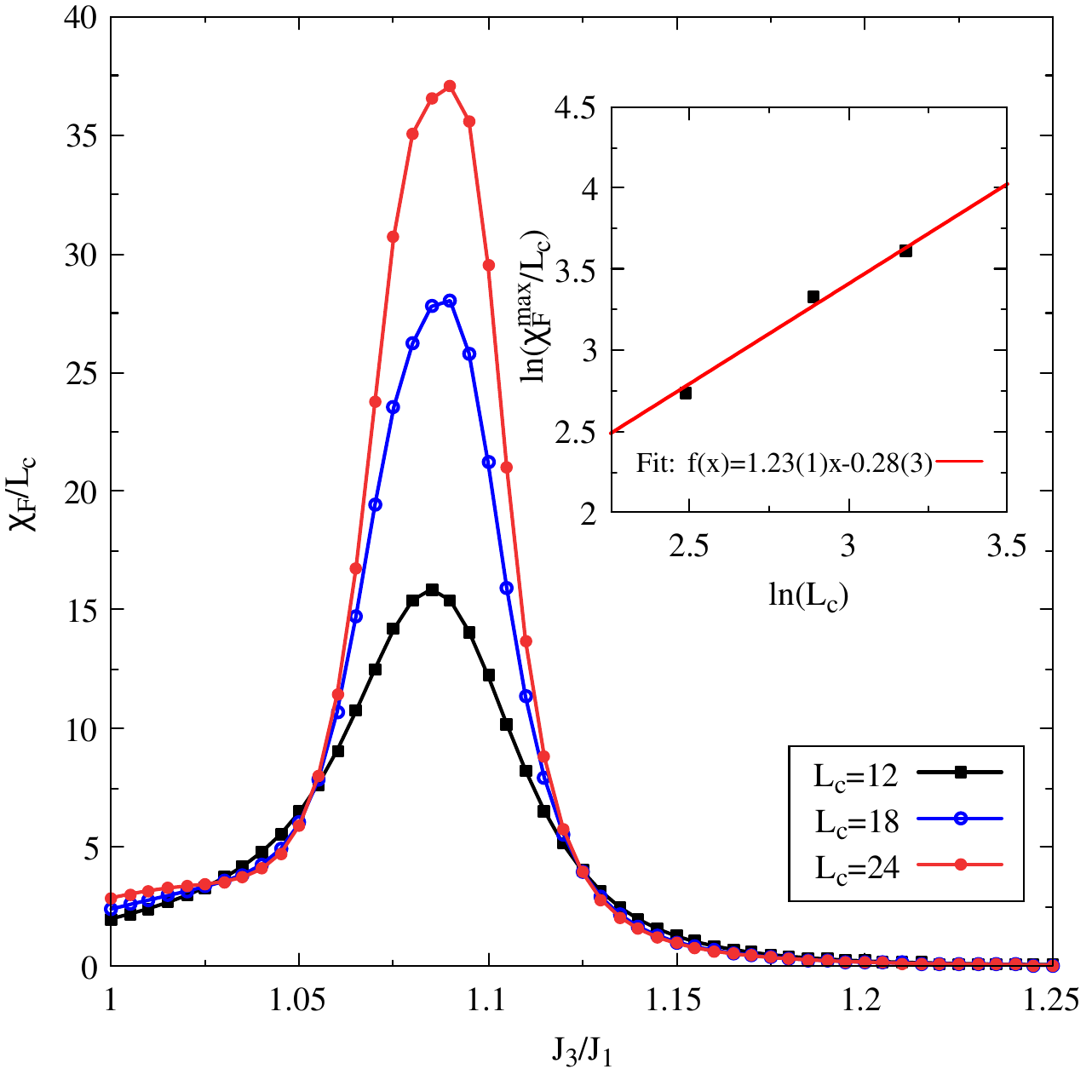}
    \caption{Fidelity susceptibility per site, $\chi_{F}/L_c$, as a function of the frustration parameter $J_3/J_1$, computed at zero magnetic field for clusters of different sizes and intertrimer coupling $J_2/J_1=0.18$. The pronounced peak structure signals enhanced ground-state sensitivity and provides a diagnostic of the quantum critical point. The inset shows the finite-size behavior of $\chi_{F}^{max}/L_c$ as a function of the system size.}
    \label{fig:fidelity}
\end{figure}

To further characterize the dynamics of the system, we also analyze the finite-size scaling of the energy gap in Fig.~\ref{fig:gs_gap}(b), expressed by $\ln(E_1-E_0)$ as a function of $\ln(L_c)$ up to $L_c=24$. 
The log-log fits yield a dynamic exponent of $z = 0.94$ for both $J_3/J_1 = 0.18$ and $J_3/J_1 = 0.50$. 
This algebraic scaling, consistent with $z \approx 1$, is a hallmark of the gapless critical phase associated with a Tomonaga-Luttinger liquid, indicating a linear energy-momentum dispersion and scale invariance \cite{giamarchi2003quantum,FDMHaldane_1981}. 
In contrast, at $J_3/J_1 = 1.00$, the scaling behavior undergoes a significant modification, yielding a dynamical critical exponent of $z \approx 2.06$. 
This value indicates the emergence of a quadratic dispersion within this crossover regime, where the finite-size energy gap decays more rapidly ($\propto 1/L^2$) as the system approaches the thermodynamic limit.
Together with the correlation function analyses, these results indicate that the instability near $J_3/J_1 \approx 1$ marks the onset of a continuous quantum phase transition.
This process, characterized by a shift towards quadratic dispersion, will culminate in a divergence in fidelity susceptibility and an enhancement of von Neumann entanglement entropy.

The fidelity susceptibility, $\chi_{F}$, can bring insights to characterize zero-field behaviors. 
By introducing small variations in the exchange interaction $J_3/J_1$, we evaluate the $\chi_{F}$ of the trimerized quantum spin chain, presenting the results in Fig.~\ref{fig:fidelity}. 
The $\chi_{F}/L_c$ dependence on the frustration parameter exhibits a pronounced peak near $J_3^{c}/J_1 \approx 1.09$ for the different cluster sizes ($L_{c}=12, 18$, and $24$ sites). 
This peak intensifies with increasing $L_{c}$, providing a signature of a quantum critical point. 
According to finite-size scaling theory, the maximum fidelity susceptibility at the critical point scales as $\chi_{F}^{\max} \sim L_{c}^{2/(d\nu)}$ \cite{GU2008,Albuquerque2010,Wang2015,Ren_2015}, where $\nu$ is the correlation length exponent and $d=1$ is the spatial dimension of the system. 
Consequently, the intensive quantity $\chi_{F}^{\max}/L_{c}$ is expected to scale as $L_c^{2/\nu - 1}$. 
This behavior is corroborated in the inset of Fig.~\ref{fig:fidelity}, where the  $\ln(\chi_{F}^{\max}/L_{c})$ is plotted as a function of $\ln(L_{c})$. 
The data display an excellent linear trend with a slope of $\alpha = 1.23(1)$. Using the relation $\alpha = 2/\nu - 1$, we extract a correlation length exponent of $\nu = 0.90(1)$. 
These results establish a scenario where increasing frustration $J_3/J_1$ drives the system from a stable Luttinger liquid phase ($z \approx 1$) to a collapse of linear dispersion at $J_3/J_1=1.0$ (where $z \approx 2$), serving as a dynamical precursor to the quantum critical point at $J_3^{c}/J_1\approx1.09$ ($\nu \approx 0.9$).

\begin{figure}[!t]
    \centering
    \includegraphics[width=0.9\linewidth]{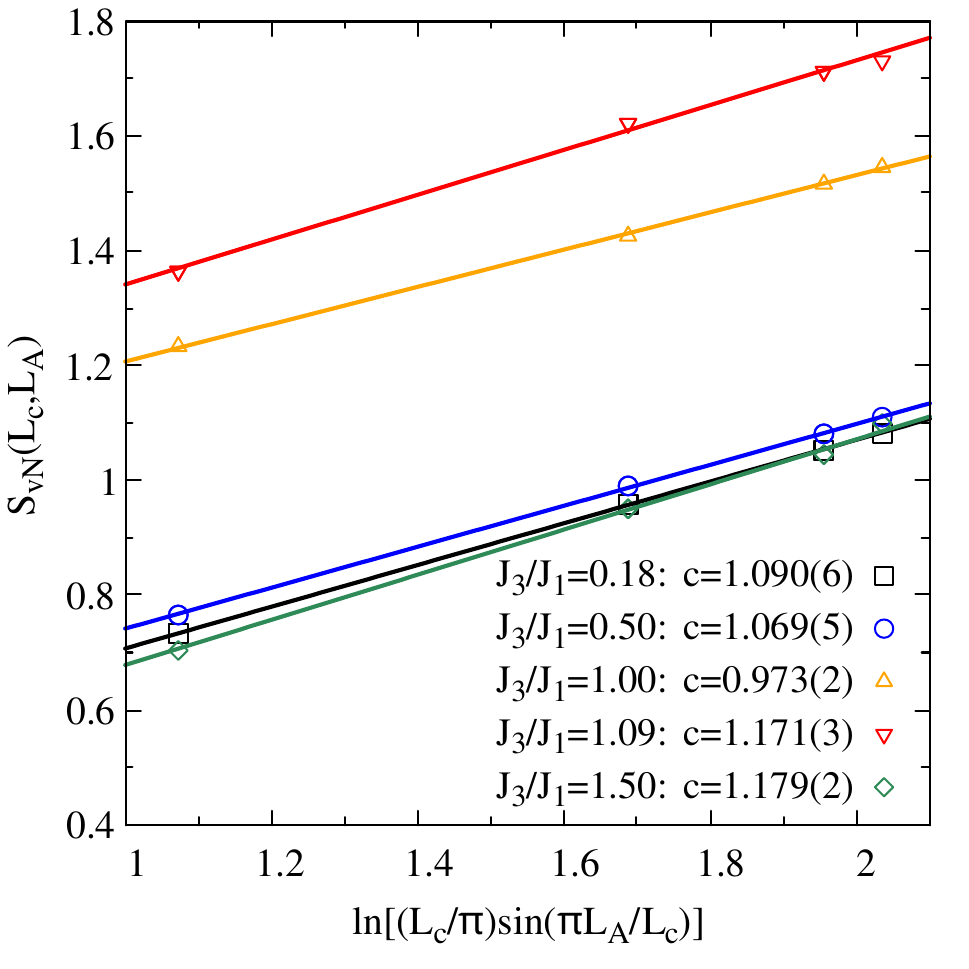}
    \caption{Von Neumann entanglement entropy $S_{vN}(L_c,L_A)$ as a function of the chord distance $\ln[(L_{c}/\pi)\sin((\pi L_A)/L_{c})]$, computed within the Lanczos diagonalization for $L_c=24$ sites and different values of frustration parameter $J_3/J_1=0.18$ (black curve), $J_3/J_1=0.50$ (blue curve), $J_3/J_1=1.00$ (orange curve), $J_3/J_1=1.09$ (red curve) and $J_3/J_1=1.50$ (green curve). The periodic boundary conditions were imposed, and the solid lines represent the logarithmic fit based on conformal field theory (see Eq. (\ref{eq:vn_scaling})).}
    \label{fig:entropy}
\end{figure}
To probe the universality class of the gapless phase identified in the preceding discussion, we calculated the von Neumann entanglement entropy $S_{vN}(L_c,L_A)$ as a function of the chord distance $\ln[(L_{c}/\pi)\sin((\pi L_A)/L_{c})]$ for different values of frustration parameter $J_3/J_1$, as shown in Fig. \ref{fig:entropy}. 
In particular, the bipartition size to evaluate the $S_{\nu N}$ was chosen as $L_
A=3l$ (with $l=1,2,...,8$), preserving the unit-cell structure across the entanglement cut.
The entropy exhibits a linear trend as a function of the chord distance. 
Linear regressions of the data reveal an excellent agreement with an effective central charge close to unity.
Specifically, we extract $c = 1.090(6)$ for $J_3/J_1 = 0.18$ and $c = 1.069(5)$ for $J_3/J_1 = 0.50$ in the low frustration regime.
These results provide evidence that the gapless low frustration region is predominantly governed by a $c \approx 1$ universality class, as expected for a Tomonaga-Luttinger liquid.
Furthermore, this behavior is consistent with CFT predictions for the spin-$1/2$ Heisenberg model in one-dimensional quantum critical systems \cite{ALCARAZ1988280,Calabrese2010,Xavier2011,Ramos2024} 
and it corroborates the DMRG calculations recently reported by Cheng \textit{et~al.}~\cite{cheng2024quantum}.

Near $J_3/J_1 = 1.00$, where finite-size scaling of the energy gap indicates a crossover toward a low-energy quadratic dispersion with $z \approx 2$, the entropy fit still returns an effective value close to unity ($c = 0.97(2)$), suggesting that logarithmic entanglement scaling may still survive, although significant finite-size corrections are expected. 
Upon increasing the frustration ratio to $J_3/J_1 = 1.09$, the entanglement entropy displays an upward shift, while the fitted effective central charge increases to $c = 1.171(3)$. This enhancement is consistent with the reorganization of the underlying correlations as the system approaches the critical point. 
On the other hand, the entanglement entropy for $J_3/J_1 > 1.09$ (see the green line for $J_3/J_1=1.50$) behaves similarly 
to that presented in the low-frustration regime ($J_3/J_1=0.18$ and 0.50).

\section{SUMMARY AND CONCLUSION}\label{sec:summary}
In summary, we have investigated the ground-state properties of a frustrated trimerized quantum spin chain employing a cluster mean-field approach combined with Lanczos exact diagonalization. 
Our results suggest that the frustration-driven reorganization of local singlet correlations provides a static microscopic precursor to the composite doublon excitations previously identified in dynamical probes of trimerized spin \cite{bera2022emergent}.

In the weakly frustrated regime, the model exhibits a robust $1/3$ magnetization plateau associated with a collective ferrimagnetic-like configuration within each trimer. 
This plateau phase is characterized by highly entangled intratrimer correlations and strongly suppressed intertrimer fluctuations, yielding an effectively local magnetic structure. 
The calculated critical fields bounding the plateau and saturation phases are in excellent quantitative agreement with previous numerical studies \cite{cheng2024quantum, kumar2025theoretical}, reinforcing the reliability of our methodology. 
As frustration increases, the evolution of local magnetizations and correlation functions indicates a reorganization of the ground state, driven primarily by the competition between the intratrimer exchange interactions $J_1$ and $J_3$.

Entering the strongly frustrated regime, the system achieves nearly perfect singlet correlations alongside a suppression of edge magnetizations. 
This behavior naturally supports a low-energy description based on composite degrees of freedom, wherein frustration-induced doublons emerge as the static structures of the ground state.
Furthermore, finite-size scaling and quantum-information diagnostics offer clear evidence of the underlying critical behavior in this regime.
The fidelity susceptibility exhibits pronounced finite-size anomalies near $J_3/J_1 \approx 1.09$, indicating a quantum critical point. 
Consistently, the von Neumann entanglement entropy displays a logarithmic scaling with subsystem size, yielding an effective central charge of $c \approx 1$. 
Remarkably, in this regime, the dynamical critical exponent changes from $z\approx1$ to an anomalous value $z\approx2$ near the transition. In this case, the relation between the central charge and $z$ suggests that the zero-field criticality evolves beyond the conventional Tomonaga–Luttinger liquid description in the strongly frustrated regime $J_{3}\rightarrow J_{1}$.

Overall, our findings reveal that frustration not only enriches the phase diagram but also fundamentally generates qualitatively distinct quantum states whose correlation structure is consistent with the emergence of nontrivial fractionalized excitations.
These insights provide a robust physical picture for understanding the unconventional magnetic behavior of trimerized quantum spin systems, offering direct theoretical relevance to the complex phenomenology observed in experimentally realized compounds such as $\mathrm{Na_2Cu_3Ge_4O_{12}}$.

A compelling open question concerns the microscopic mechanism through which quarton-like excitations, which are interpreted as higher-order composite modes. 
Such excitations are expected to reflect the collective nature of correlations in the strongly frustrated regime, establishing a direct connection between local singlet formation and emergent multi-spin bound states. A detailed investigation of this mechanism is beyond the present CMF method and remains a highly promising direction for future research.

\section{ACKNOWLEDGMENTS}\label{sec:acknowledgment}
We thank Francisco C. Alcaraz for the discussions. 
This work was supported by Brazilian agencies Conselho Nacional de Desenvolvimento Científico e Tecnológico (CNPq), processes 165330/2023-6 (FMZ) and 309652/2023-5 (MS), and Coordenação de Aperfeiçoamento de Pessoal de Nível Superior (Capes). 
FMZ and LMR also acknowledge support from the Fundação de Apoio ao Desenvolvimento do Ensino, Ciência e Tecnologia do Estado de Mato Grosso do Sul (Fundect).

\bibliography{bib}

@misc{sen2026trimerizedspin12chainemergent,
      title={Trimerized Spin-$1/2$ Chain: Emergent Low-Energy Hamiltonian, Higher-Energy Excitations, and Magnetic and Thermodynamic Responses}, 
      author={Snehasish Sen and Sudhansu S. Mandal},
      year={2026},
      eprint={2605.25960},
      archivePrefix={arXiv},
      primaryClass={cond-mat.str-el},
}

@article{lanczos1950iteration,
  title={An iteration method for the solution of the eigenvalue problem of linear differential and integral operators},
  author={Lanczos, Cornelius},
  journal={Journal of research of the National Bureau of Standards},
  volume={45},
  number={4},
  pages={255--282},
  year={1950},
  url={https://doi.org/10.6028/jres.045.026}
}

@book{giamarchi2003quantum,
    author = {Giamarchi, Thierry},
    title = {Quantum Physics in One Dimension},
    publisher = {Oxford University Press},
    year = {2003},
    month = {12},
    isbn = {9780198525004},
    doi = {10.1093/acprof:oso/9780198525004.001.0001},
    url = {https://doi.org/10.1093/acprof:oso/9780198525004.001.0001},
}

@article{jordan1928paulische,
  title={{\"U}ber das paulische {\"a}quivalenzverbot},
  author={Jordan, Pascual and Wigner, Eugene},
  journal={Zeitschrift f{\"u}r Physik},
  volume={47},
  number={9},
  pages={631--651},
  year={1928},
  publisher={Springer},
  url={https://doi.org/10.1007/BF01331938},
  doi={10.1007/BF01331938}
}

@article{mikeska2004one,
  title={One-Dimensional Magnetism in Quantum Magnetism, ed. U.},
  author={Mikeska, HJ and Kolezhuk, AK},
  journal={Lecture Notes in Physics},
  volume={645},
  url={https://doi.org/10.1007/BFb0119591},
  year={2004}
}

@article{hellman1937einfuhrung,
  title={Einf{\"u}hrung in die Quantenchemie},
  author={Hellman, Hans},
  journal={Franz Deuticke, Leipzig},
  volume={285},
  pages={90},
  year={1937}
}

@article{Feynman1939,
  title = {Forces in Molecules},
  author = {Feynman, R. P.},
  journal = {Phys. Rev.},
  volume = {56},
  issue = {4},
  pages = {340--343},
  numpages = {0},
  year = {1939},
  month = {Aug},
  publisher = {American Physical Society},
  doi = {10.1103/PhysRev.56.340},
  url = {https://link.aps.org/doi/10.1103/PhysRev.56.340}
}

@book{parlett1998symmetric,
  author = {Parlett, Beresford N.},
title = {The Symmetric Eigenvalue Problem},
publisher = {Society for Industrial and Applied Mathematics},
year = {1998},
doi = {10.1137/1.9781611971163},
address = {},
edition   = {},
URL = {https://epubs.siam.org/doi/abs/10.1137/1.9781611971163},
}

@article{bera2022emergent,
  title={Emergent many-body composite excitations of interacting spin-1/2 trimers},
  author={Bera, Anup Kumar and Yusuf, SM and Saha, Sudip Kumar and Kumar, Manoranjan and Voneshen, David and Skourski, Yurii and Zvyagin, Sergei A},
  journal={Nature Communications},
  volume={13},
  number={1},
  pages={6888},
  year={2022},
  publisher={Nature Publishing Group UK London},
  url={https://doi.org/10.1038/s41467-022-34342-1},
  doi={10.1038/s41467-022-34342-1}
}

@article{cheng2024quantum,
  title={Quantum phase transition and composite excitations of antiferromagnetic spin trimer chains in a magnetic field},
  author={Cheng, Jun-Qing and Ning, Zhi-Yao and Wu, Han-Qing and Yao, Dao-Xin},
  journal={npj Quantum Materials},
  volume={9},
  number={1},
  pages={96},
  year={2024},
  publisher={Nature Publishing Group UK London},
  url={https://doi.org/10.1038/s41535-024-00705-8},
  doi={10.1038/s41535-024-00705-8}
}

@article{GuBO2006,
  title = {Thermodynamics of spin-$1/2$ antiferromagnet-antiferromagnet-ferromagnet and ferromagnet-ferromagnet-antiferromagnet trimerized quantum Heisenberg chains},
  author = {Gu, Bo and Su, Gang and Gao, Song},
  journal = {Phys. Rev. B},
  volume = {73},
  issue = {13},
  pages = {134427},
  numpages = {10},
  year = {2006},
  month = {Apr},
  publisher = {American Physical Society},
  doi = {10.1103/PhysRevB.73.134427},
  url = {https://link.aps.org/doi/10.1103/PhysRevB.73.134427}
}

@article{okamoto1999magnetization,
 doi = {10.1088/0305-4470/32/25/304},
url = {https://doi.org/10.1088/0305-4470/32/25/304},
year = {1999},
month = {jun},
publisher = {},
volume = {32},
number = {25},
pages = {4601},
author = {Kiyomi Okamoto and Atsuhiro Kitazawa},
title = {Magnetization plateau and quantum phase transition of the {$S = 1/2$} trimerized XXZ spin chain},
journal = {Journal of Physics A: Mathematical and General}
}

@article{Verkholyak2021,
  title = {Modified strong-coupling treatment of a spin-$\frac{1}{2}$ Heisenberg trimerized chain developed from the exactly solved Ising-Heisenberg diamond chain},
  author = {Verkholyak, Taras and Stre\ifmmode \check{c}\else \v{c}\fi{}ka, Jozef},
  journal = {Phys. Rev. B},
  volume = {103},
  issue = {18},
  pages = {184415},
  numpages = {10},
  year = {2021},
  month = {May},
  publisher = {American Physical Society},
  doi = {10.1103/PhysRevB.103.184415},
  url = {https://link.aps.org/doi/10.1103/PhysRevB.103.184415}
}

@article{oguchi1955theory,
 author = {Oguchi, Takehiko},
    title = {A Theory of Antiferromagnetism, II},
    journal = {Progress of Theoretical Physics},
    volume = {13},
    number = {2},
    pages = {148-159},
    year = {1955},
    month = {02},
    issn = {0033-068X},
    doi = {10.1143/PTP.13.148},
    url = {https://doi.org/10.1143/PTP.13.148},
}

@article{schmidt2017spin,
doi = {10.1088/1361-648X/aa6060},
url = {https://dx.doi.org/10.1088/1361-648X/aa6060},
year = {2017},
month = {mar},
publisher = {IOP Publishing},
volume = {29},
number = {16},
pages = {165801},
author = {Schmidt, M and Zimmer, F M and Magalhaes, S G},
title = {Spin liquid and infinitesimal-disorder-driven cluster spin glass in the kagome lattice},
journal = {Journal of Physics: Condensed Matter}
}

@article{malakar2020phases,
 title = {Phases and collective modes of bosons in a triangular lattice at finite temperature: A cluster mean field study},
  author = {Malakar, M. and Ray, S. and Sinha, S. and Angom, D.},
  journal = {Phys. Rev. B},
  volume = {102},
  issue = {18},
  pages = {184515},
  numpages = {11},
  year = {2020},
  month = {Nov},
  publisher = {American Physical Society},
  doi = {10.1103/PhysRevB.102.184515},
  url = {https://link.aps.org/doi/10.1103/PhysRevB.102.184515}
}

@article{zimmer2016quantum,
  title = {Quantum correlated cluster mean-field theory applied to the transverse Ising model},
  author = {Zimmer, F. M. and Schmidt, M. and Maziero, Jonas},
  journal = {Phys. Rev. E},
  volume = {93},
  issue = {6},
  pages = {062116},
  numpages = {8},
  year = {2016},
  month = {Jun},
  publisher = {American Physical Society},
  doi = {10.1103/PhysRevE.93.062116},
  url = {https://link.aps.org/doi/10.1103/PhysRevE.93.062116}
}

@article{kellermann2019quantum,
  title = {Quantum Ising model on the frustrated square lattice},
  author = {Kellermann, N. and Schmidt, M. and Zimmer, F. M.},
  journal = {Phys. Rev. E},
  volume = {99},
  issue = {1},
  pages = {012134},
  numpages = {6},
  year = {2019},
  month = {Jan},
  publisher = {American Physical Society},
  doi = {10.1103/PhysRevE.99.012134},
  url = {https://link.aps.org/doi/10.1103/PhysRevE.99.012134}
}

@article{godoy2020ising,
  title = {The Ising model on the layered {$J_1-J_2$} square lattice},
journal = {Physics Letters A},
volume = {384},
number = {27},
pages = {126687},
year = {2020},
issn = {0375-9601},
doi = {https://doi.org/10.1016/j.physleta.2020.126687},
url = {https://www.sciencedirect.com/science/article/pii/S0375960120305545},
author = {P. {F. Godoy} and M. Schmidt and F. {M. Zimmer}}
}

@article{jin2013phase,
   title = {Phase transitions in the frustrated Ising model on the square lattice},
  author = {Jin, Songbo and Sen, Arnab and Guo, Wenan and Sandvik, Anders W.},
  journal = {Phys. Rev. B},
  volume = {87},
  issue = {14},
  pages = {144406},
  numpages = {12},
  year = {2013},
  month = {Apr},
  publisher = {American Physical Society},
  doi = {10.1103/PhysRevB.87.144406},
  url = {https://link.aps.org/doi/10.1103/PhysRevB.87.144406}
}

@article{WIESER2021168414,
title = {Cluster mean-field theory studies of the frustrated two-dimensional quantum-mechanical {$J_1-J_2$ Heisenberg} model},
journal = {Annals of Physics},
volume = {427},
pages = {168414},
year = {2021},
issn = {0003-4916},
doi = {https://doi.org/10.1016/j.aop.2021.168414},
url = {https://www.sciencedirect.com/science/article/pii/S0003491621000208},
author = {R. Wieser}
}

@article{PhysRevResearch.2.043013,
  title = {Intrinsic quantum {I}sing model on a triangular lattice magnet $\mathrm{Tm}\mathrm{Mg}\mathrm{Ga}${O}$_{4}$},
  author = {Liu, Changle and Huang, Chun-Jiong and Chen, Gang},
  journal = {Phys. Rev. Res.},
  volume = {2},
  issue = {4},
  pages = {043013},
  numpages = {15},
  year = {2020},
  month = {Oct},
  publisher = {American Physical Society},
  doi = {10.1103/PhysRevResearch.2.043013},
  url = {https://link.aps.org/doi/10.1103/PhysRevResearch.2.043013}
}

@article{Singhania,
  title = {Cluster mean-field study of the {Heisenberg} model for {${\mathrm{CuInVO}}_{5}$}},
  author = {Singhania, Ayushi and Kumar, Sanjeev},
  journal = {Phys. Rev. B},
  volume = {98},
  issue = {10},
  pages = {104429},
  numpages = {9},
  year = {2018},
  month = {Sep},
  publisher = {American Physical Society},
  doi = {10.1103/PhysRevB.98.104429},
  url = {https://link.aps.org/doi/10.1103/PhysRevB.98.104429}
}

@article{ramos2025interplay,
   title = {Interplay of frustration and quantum fluctuations in a spin-1/2 anisotropic square lattice},
  author = {Ramos, L. M. and Zimmer, F. M. and Schmidt, M.},
  journal = {Phys. Rev. B},
  volume = {112},
  issue = {1},
  pages = {014402},
  numpages = {9},
  year = {2025},
  month = {Jul},
  publisher = {American Physical Society},
  doi = {10.1103/svsc-bvjx},
  url = {https://link.aps.org/doi/10.1103/svsc-bvjx}
}

@article{RAMOS2026173687,
title = {Spin–singlet dimer phase in a frustrated square lattice under a magnetic field},
journal = {Journal of Magnetism and Magnetic Materials},
volume = {637},
pages = {173687},
year = {2026},
issn = {0304-8853},
doi = {https://doi.org/10.1016/j.jmmm.2025.173687},
url = {https://www.sciencedirect.com/science/article/pii/S0304885325009217},
author = {L.M. Ramos and M. Schmidt and F.M. Zimmer}
}

@article{zanardi2006,
  title = {Ground state overlap and quantum phase transitions},
  author = {Zanardi, Paolo and Paunkovi\ifmmode \acute{c}\else \'{c}\fi{}, Nikola},
  journal = {Phys. Rev. E},
  volume = {74},
  issue = {3},
  pages = {031123},
  numpages = {6},
  year = {2006},
  month = {Sep},
  publisher = {American Physical Society},
  doi = {10.1103/PhysRevE.74.031123},
  url = {https://link.aps.org/doi/10.1103/PhysRevE.74.031123}
}

@article{PhysRevLett.114.027201,
  title = {Microscopic Model Calculations for the Magnetization Process of Layered Triangular-Lattice Quantum Antiferromagnets},
  author = {Yamamoto, Daisuke and Marmorini, Giacomo and Danshita, Ippei},
  journal = {Phys. Rev. Lett.},
  volume = {114},
  issue = {2},
  pages = {027201},
  numpages = {5},
  year = {2015},
  month = {Jan},
  publisher = {American Physical Society},
  doi = {10.1103/PhysRevLett.114.027201}
}

@article{yamamoto2009,
  title = {Correlated cluster mean-field theory for spin systems},
  author = {Yamamoto, Daisuke},
  journal = {Phys. Rev. B},
  volume = {79},
  issue = {14},
  pages = {144427},
  numpages = {9},
  year = {2009},
  month = {Apr},
  publisher = {American Physical Society},
  doi = {10.1103/PhysRevB.79.144427},
  url = {https://link.aps.org/doi/10.1103/PhysRevB.79.144427}
}

@misc{patnaik2025fermionic,
title={Fermionic Band Dispersions and an Evidence of Cooperon Excitations in a Spin-$1/2$ Trimer Chain}, 
author={P. Srikanth Patnaik and Snehasish Sen and A. K. Bera and Sudhansu S. Mandal and Anushree Roy and S. M. Yusuf},
year={2025},
eprint={2504.21616},
archivePrefix={arXiv},
primaryClass={cond-mat.str-el},
}

@article{prabhakar2025fractionalized,
  title = {Fractionalized excitations and resonant inelastic x-ray spectra in frustrated spin-1/2 trimer chains},
  author = {Prabhakar and Pal, Subhajyoti and Kumar, Umesh and Kumar, Manoranjan and Mukherjee, Anamitra},
  journal = {Phys. Rev. B},
  volume = {111},
  issue = {20},
  pages = {205106},
  numpages = {12},
  year = {2025},
  month = {May},
  publisher = {American Physical Society},
  doi = {10.1103/PhysRevB.111.205106},
  url = {https://link.aps.org/doi/10.1103/PhysRevB.111.205106}
}

@article{li2025resonant,
 title = {Resonant inelastic x-ray scattering spectra of spinon, doublon, and quarton excitations of a spin-$\frac{1}{2}$ antiferromagnetic Heisenberg trimer chain},
  author = {Li, Junli and Cheng, Jun-Qing and Datta, Trinanjan and Yao, Dao-Xin},
  journal = {Phys. Rev. B},
  volume = {111},
  issue = {2},
  pages = {024404},
  numpages = {13},
  year = {2025},
  month = {Jan},
  publisher = {American Physical Society},
  doi = {10.1103/PhysRevB.111.024404},
  url = {https://link.aps.org/doi/10.1103/PhysRevB.111.024404}
}

@article{chikara2023role,
  author = {Chikara, Kuldeep Singh and Bera, Anup Kumar and Kumar, Amit and Yusuf, Seikh Mohammad},
title = {Role of Crystal Structure on the Ionic Conduction and Electrical Properties of Germanate Compounds {$\mathrm{A}_2 \mathrm{Cu}_3 \mathrm{Ge}_4 \mathrm{O}_{12}$} ($\mathrm{A}$ = $\mathrm{Na}$, $\mathrm{K}$)},
journal = {ACS Applied Electronic Materials},
volume = {5},
number = {5},
pages = {2704-2717},
year = {2023},
doi = {10.1021/acsaelm.3c00176},
URL = {https://doi.org/10.1021/acsaelm.3c00176}
}

@article{hida1994magnetic,
author = {Hida ,Kazuo},
title = {Magnetic Properties of the Spin-1/2    Ferromagnetic-Ferromagnetic-Antiferromagnetic    Trimerized Heisenberg Chain},
journal = {Journal of the Physical Society of Japan},
volume = {63},
number = {6},
pages = {2359-2364},
year = {1994},
doi = {10.1143/JPSJ.63.2359},
URL = {https://doi.org/10.1143/JPSJ.63.2359}
}

@article{savary2017quantum,
  doi = {10.1088/0034-4885/80/1/016502},
url = {https://doi.org/10.1088/0034-4885/80/1/016502},
year = {2016},
month = {nov},
publisher = {IOP Publishing},
volume = {80},
number = {1},
pages = {016502},
author = {Savary, Lucile and Balents, Leon},
title = {Quantum spin liquids: a review},
journal = {Reports on Progress in Physics}
}

@article{balents2010spin,
  title={Spin liquids in frustrated magnets},
  author={Balents, Leon},
  journal={Nature},
  volume={464},
  number={7286},
  pages={199--208},
  year={2010},
  publisher={Nature Publishing Group UK London},
  url={https://doi.org/10.1038/nature08917},
  doi={10.1038/nature08917}
}

@article{anderson1987resonating,
author = {P. W. Anderson },
title = {The Resonating Valence Bond State in {$\mathrm{La}_2\mathrm{CuO}_4$} and Superconductivity},
journal = {Science},
volume = {235},
number = {4793},
pages = {1196-1198},
year = {1987},
doi = {10.1126/science.235.4793.1196},
URL = {https://www.science.org/doi/abs/10.1126/science.235.4793.1196}}

@article{lieb1961two,
title = {Two soluble models of an antiferromagnetic chain},
journal = {Annals of Physics},
volume = {16},
number = {3},
pages = {407-466},
year = {1961},
issn = {0003-4916},
doi = {https://doi.org/10.1016/0003-4916(61)90115-4},
url = {https://www.sciencedirect.com/science/article/pii/0003491661901154},
author = {Elliott Lieb and Theodore Schultz and Daniel Mattis}
}

@article{yasui2014magnetic,
 author = {Yasui, Yukio and Kawamura, Yuji and Kobayashi, Yoshiaki and Sato, Masatoshi},
    title = {Magnetic and dielectric properties of one-dimensional array of S=1/2 linear trimer system {$\mathrm{Na}_2\mathrm{Cu}_3\mathrm{Ge}_4\mathrm{O}_{12}$}},
    journal = {Journal of Applied Physics},
    volume = {115},
    number = {17},
    pages = {17E125},
    year = {2014},
    month = {02},
    issn = {0021-8979},
    doi = {10.1063/1.4865776},
    url = {https://doi.org/10.1063/1.4865776}
}

@article{Han2024,
  title = {High-field NMR study on the topologically nontrivial 1/3 magnetization plateau state in doped {${\mathrm{Na}}_{2}{\mathrm{Cu}}_{3}{\mathrm{Ge}}_{4\ensuremath{-}x}{\mathrm{Si}}_{x}{\mathrm{O}}_{12}$}},
  author = {Han, Yuyan and Yu, Bocheng and Du, Zan and Ling, Langsheng and Zhang, Lei and Tong, Wei and Xi, Chuanying and Zhang, Jinglei and Shang, Tian and Pi, Li and Ma, Long},
  journal = {Phys. Rev. B},
  volume = {110},
  issue = {20},
  pages = {L201102},
  numpages = {6},
  year = {2024},
  month = {Nov},
  publisher = {American Physical Society},
  doi = {10.1103/PhysRevB.110.L201102},
  url = {https://link.aps.org/doi/10.1103/PhysRevB.110.L201102}
}

@article{kumar2025theoretical,
  title={Theoretical and experimental studies of melting of the 1/3 magnetization plateau in a frustrated S= 1/2 antiferromagnetic trimerized quantum Heisenberg spin chain compound {$\mathrm{Na}_2\mathrm{Cu}_3\mathrm{Ge}_4\mathrm{O}_{12}$}},
  author={Kumar, Sachin and Bera, AK and Kumar, Amit and Skourski, Yurii and Yusuf, SM},
  journal={The European Physical Journal B},
  volume={98},
  number={3},
  pages={46},
  year={2025},
  publisher={Springer},
  url={https://doi.org/10.1140/epjb/s10051-025-00891-9},
  doi={10.1140/epjb/s10051-025-00891-9}
}

@article{bethe1931theorie,
  title={Zur theorie der metalle: I. Eigenwerte und eigenfunktionen der linearen atomkette},
  author={Bethe, Hans},
  journal={Zeitschrift f{\"u}r Physik},
  volume={71},
  number={3},
  pages={205--226},
  year={1931},
  publisher={Springer},
  url ={https://doi.org/10.1007/BF01341708},
  doi={10.1007/BF01341708}
}

@article{Calabrese2010,
  title = {Parity Effects in the Scaling of Block Entanglement in Gapless Spin Chains},
  author = {Calabrese, Pasquale and Campostrini, Massimo and Essler, Fabian and Nienhuis, Bernard},
  journal = {Phys. Rev. Lett.},
  volume = {104},
  issue = {9},
  pages = {095701},
  numpages = {4},
  year = {2010},
  month = {Mar},
  publisher = {American Physical Society},
  doi = {10.1103/PhysRevLett.104.095701},
  url = {https://link.aps.org/doi/10.1103/PhysRevLett.104.095701}
}

@article{ALCARAZ1988280,
title = {Conformal invariance, the XXZ chain and the operator content of two-dimensional critical systems},
journal = {Annals of Physics},
volume = {182},
number = {2},
pages = {280-343},
year = {1988},
issn = {0003-4916},
doi = {https://doi.org/10.1016/0003-4916(88)90015-2},
url = {https://www.sciencedirect.com/science/article/pii/0003491688900152},
author = {Francisco C Alcaraz and Michael N Barber and Murray T Batchelor}
}

@article{calabrese2009,
doi = {10.1088/1751-8113/42/50/504005},
url = {https://doi.org/10.1088/1751-8113/42/50/504005},
year = {2009},
month = {dec},
publisher = {},
volume = {42},
number = {50},
pages = {504005},
author = {Calabrese, Pasquale and Cardy, John},
title = {Entanglement entropy and conformal field theory},
journal = {Journal of Physics A: Mathematical and Theoretical}
}

@article{FDMHaldane_1981,
doi = {10.1088/0022-3719/14/19/010},
url = {https://doi.org/10.1088/0022-3719/14/19/010},
year = {1981},
month = {jul},
publisher = {},
volume = {14},
number = {19},
pages = {2585},
author = {F D M Haldane},
title = {Luttinger liquid theory of one-dimensional quantum fluids. I. Properties of the Luttinger model and their extension to the general 1D interacting spinless Fermi gas},
journal = {Journal of Physics C: Solid State Physics}
}

@article{Ren_2015,
doi = {10.1088/0953-8984/27/10/105602},
url = {https://doi.org/10.1088/0953-8984/27/10/105602},
year = {2015},
month = {feb},
publisher = {IOP Publishing},
volume = {27},
number = {10},
pages = {105602},
author = {Ren, Jie and Liu, Guang-Hua and You, Wen-Long},
title = {Entanglement entropy and fidelity susceptibility in the one-dimensional spin-1 XXZ chains with alternating single-site anisotropy},
journal = {Journal of Physics: Condensed Matter}
}

@article{Albuquerque2010,
  title = {Quantum critical scaling of fidelity susceptibility},
  author = {Albuquerque, A. Fabricio and Alet, Fabien and Sire, Cl\'ement and Capponi, Sylvain},
  journal = {Phys. Rev. B},
  volume = {81},
  issue = {6},
  pages = {064418},
  numpages = {12},
  year = {2010},
  month = {Feb},
  publisher = {American Physical Society},
  doi = {10.1103/PhysRevB.81.064418},
  url = {https://link.aps.org/doi/10.1103/PhysRevB.81.064418}
}

@article{GU2008,
  title = {Fidelity susceptibility, scaling, and universality in quantum critical phenomena},
  author = {Gu, Shi-Jian and Kwok, Ho-Man and Ning, Wen-Qiang and Lin, Hai-Qing},
  journal = {Phys. Rev. B},
  volume = {77},
  issue = {24},
  pages = {245109},
  numpages = {5},
  year = {2008},
  month = {Jun},
  publisher = {American Physical Society},
  doi = {10.1103/PhysRevB.77.245109},
  url = {https://link.aps.org/doi/10.1103/PhysRevB.77.245109}
}

@article{Wang2015,
  title = {Fidelity Susceptibility Made Simple: A Unified Quantum Monte Carlo Approach},
  author = {Wang, Lei and Liu, Ye-Hua and Imri\ifmmode \check{s}\else \v{s}\fi{}ka, Jakub and Ma, Ping Nang and Troyer, Matthias},
  journal = {Phys. Rev. X},
  volume = {5},
  issue = {3},
  pages = {031007},
  numpages = {14},
  year = {2015},
  month = {Jul},
  publisher = {American Physical Society},
  doi = {10.1103/PhysRevX.5.031007},
  url = {https://link.aps.org/doi/10.1103/PhysRevX.5.031007}
}

@article{Ramos2024,
  title = {Conformally invariant free-parafermionic quantum chains with multispin interactions},
  author = {Alcaraz, Francisco C. and Ramos, Lucas M.},
  journal = {Phys. Rev. E},
  volume = {109},
  issue = {4},
  pages = {044138},
  numpages = {15},
  year = {2024},
  month = {Apr},
  publisher = {American Physical Society},
  doi = {10.1103/PhysRevE.109.044138},
  url = {https://link.aps.org/doi/10.1103/PhysRevE.109.044138}
}

@article{Xavier2011,
  title = {R\'enyi entropy and parity oscillations of anisotropic spin-$s$ Heisenberg chains in a magnetic field},
  author = {Xavier, J. C. and Alcaraz, F. C.},
  journal = {Phys. Rev. B},
  volume = {83},
  issue = {21},
  pages = {214425},
  numpages = {8},
  year = {2011},
  month = {Jun},
  publisher = {American Physical Society},
  doi = {10.1103/PhysRevB.83.214425},
  url = {https://link.aps.org/doi/10.1103/PhysRevB.83.214425}
}

@article{CAROLLO20201,
title = {Geometry of quantum phase transitions},
journal = {Physics Reports},
volume = {838},
pages = {1-72},
year = {2020},
issn = {0370-1573},
doi = {https://doi.org/10.1016/j.physrep.2019.11.002},
url = {https://www.sciencedirect.com/science/article/pii/S0370157319303655},
author = {Angelo Carollo and Davide Valenti and Bernardo Spagnolo}
}

@article{GU2010,
author = {GU, SHI-JIAN},
title = {FIDELITY APPROACH TO QUANTUM PHASE TRANSITIONS},
journal = {International Journal of Modern Physics B},
volume = {24},
number = {23},
pages = {4371-4458},
year = {2010},
doi = {10.1142/S0217979210056335},
URL = {https://doi.org/10.1142/S0217979210056335}
}

@article{Pasquale2004,
doi = {10.1088/1742-5468/2004/06/P06002},
url = {https://doi.org/10.1088/1742-5468/2004/06/P06002},
year = {2004},
month = {jun},
publisher = {},
volume = {2004},
number = {06},
pages = {P06002},
author = {Pasquale Calabrese and John Cardy},
title = {Entanglement entropy and quantum field theory},
journal = {Journal of Statistical Mechanics: Theory and Experiment}
}

\end{document}